\documentclass[final,5p,times,twocolumn]{elsarticle}
\usepackage{amsmath,latexsym}

\usepackage{hyperref}

\journal{Physics Letters B}

\usepackage{bm}
\usepackage{graphicx,color}
\usepackage{physics}
\usepackage[normalem]{ulem}
\usepackage{hyperref}
\usepackage{amsmath}
\usepackage{float}
\usepackage{subfigure}
\usepackage{graphicx}
\usepackage{tabularx}
\usepackage{colortbl}
\usepackage{dcolumn}
\usepackage{array}

\usepackage{varioref}
\usepackage[active]{srcltx}

\expandafter\ifx\csname package@font\endcsname\relax\else
\expandafter\expandafter
\expandafter\usepackage
\expandafter\expandafter
\expandafter{\csname package@font\endcsname}%
\fi

\hypersetup{
	colorlinks=true,       
	linkcolor=blue,          
	citecolor=blue,        
}
\definecolor{myOrange}{rgb}{0.75,0.25,0.45}

\newcommand{\tcr}[1]{\textcolor{red}{#1}}
\usepackage{comment}

\usepackage[section]{placeins}

\begin{document}

\begin{frontmatter}
\title{Quantum gravitational signatures in next-generation gravitational wave detectors}
\author{Saurya Das\fnref{saurya.das@uleth.ca}}
\address{Theoretical Physics Group and Quantum Alberta, Department of Physics and Astronomy, University of Lethbridge, 4401 University Drive, Lethbridge, Alberta, T1K 3M4, Canada}
\author{S. Shankaranarayanan\fnref{shanki@phy.iitb.ac.in}}
\address{Department of Physics, Indian Institute of Technology Bombay, Mumbai 400076, India}
\author{Vasil Todorinov\fnref{Corresponding Author, v.todorinov@uleth.ca}}
 \address{Theoretical Physics Group and Quantum Alberta, Department of Physics and Astronomy, University of Lethbridge, 4401 University Drive, Lethbridge, Alberta, T1K 3M4, Canada}
\begin{abstract}
A recent study established a correspondence between the Generalized Uncertainty Principle (GUP) and Modified theories of gravity, particularly Stelle gravity.
We investigate the consequences of this correspondence for inflation and cosmological observables by evaluating the power spectrum of the scalar and tensor perturbations using two distinct methods.
First, we employ PLANCK observations to determine the GUP parameter $\gamma_0$. Then, we use the value of $\gamma_0$ to investigate the implications of quantum gravity on the power spectrum of primordial gravitational waves and their possible detectability in the next-generation detectors, like Einstein Telescope and Cosmic explorer. 
\end{abstract}

\end{frontmatter}


%
Since its discovery, the cosmic microwave background (CMB) has been an indispensable tool for understanding the very early universe. Consequently, observational cosmology has made incredible progress over the last three decades and thanks to the observation of temperature fluctuations at the last scattering surface, one has been able to identify the physical origins of the primordial density perturbations in the very early universe~\cite{COBE:1992syq,Boomerang:2000efg,WMAP:2003elm,BICEP2expt,Planck:2018nkj,Planck:2018jri,2021-BICEP3-PLANCK}. The inflationary paradigm provides a causal mechanism for the origin of these perturbations and the inflationary epoch magnifies the tiny fluctuations in the quantum fields present at the beginning of the epoch into classical perturbations that leave an imprint as anisotropies in the CMB~\cite{1980-Starobinsky-PLB,1984-Kodama.Sasaki-PTPS,1992-Mukhanov.etal-PRep,1995-Lidsey.etal-RMP,1999-Lyth.Riotto-PRep,2006-Bassett.etal-RMP}. A key prediction of inflation is 
the generation of primordial gravitational waves (PGWs)~\cite{1990-Grishchuk.Sidorov-PRD,2009-Sathyaprakash.Schutz-LRR,2016-Guzzetti.etal-Review}. 

Therefore, significant effort is currently underway to detect 
PGWs via B-mode polarization measurements of the CMB~\cite{CMBPolStudyTeam:2008rgp,Crill:2008rd,2012CLASS,BICEP2:2015nss,POLARBEAR:2014hgp,2016PIPER,Planck:2018nkj}. This is because B-modes of CMB are \emph{only} sourced by the differential stretching of spacetime associated with the PGWs~\cite{2016-Guzzetti.etal-Review,Paoletti:2022anb}. However, the primordial B-mode polarization signal is weak and could be swamped inside dust emission in our galaxy~\cite{Crowder:2005nr,BICEP2:2015nss,Planck:2018nkj}. 
Efforts are also underway to directly detect PGWs~\cite{Hild:2010id,Sato:2017dkf,2017-LISA,Maggiore:2019uih,Evans:2021gyd,Srivastava:2022slt}. In this work, we explicitly show that PGWs carry quantum gravitational signatures that can potentially be observed in the next-generation gravitational wave detectors such as the Einstein Telescope (ET) and Cosmic Explorer (CE).

Theories of Quantum Gravity (QG) predict the existence of a minimum measurable length and/or a maximum measurable momentum of particles~\cite{1994-Garay-IJMPD,2012-Hossenfelder-LRR}. The experimental implications of this scale are explored in Quantum Gravity Phenomenology (QGP)~\cite{2022-Addazi.etal-PPNP}.
This QGP/QG scale is most easily 
realized by deforming the standard Heisenberg uncertainty relation to the so-called Generalized Uncertainty Principle (GUP)~\cite{Adler1999-db,
Adler_2001,
Ali_2014,
Ali_2015,
Alonso_Serrano_2018,
Amati1989-gs,
Amelino-Camelia2013-xs,
Bargue_o_2015,
Bambi2007-te,
Bawaj_2015,
Bojowald2011-bb,
Bolen2005-jq,
bosso2017generalized,
Bosso2018,
Bosso:2018uus,
Bosso:2019ljf,
Burger_2018,
bushev2019testing,
Casadio_2020,
Chang:2011jj,
Cortes:2004qn,
Costa_Filho2016-ox,
Dabrowski2020-kk,
Das2008,
Das:2010zf,
Das_2019,
Das_2020,
Garcia-Chung:2020zyq,
Giddings2020-xz,
Hamil2019-qh,
Hossenfelder:2006cw,
Kempf1995-ka,
Kober:2010sj,
KONISHI1990276,
MAGGIORE199365,
Marin:2013pga,
Moradpour2021-jy,
Mureika2019-lf,
Myung_2007,
Park2008-uj,
Snyder:1946qz,
Sprenger_2011,
Stargen_2019,
wang2016solutions,
Bosso2020-dv,
Bosso2020-fz}. Recently, we proposed a Lorentz invariant implementation of GUP, wherein we modified the canonical commutation relation to~\cite{Todorinov2018-xi}:
 \begin{equation}
     [x^{\mu},p^{\nu}]=i\hbar\eta^{\mu\nu}(1 - \gamma p^{\sigma}p_{\sigma})-2i\hbar\gamma p^{\mu}p^{\nu},
     \label{def:GUPcomm}
 \end{equation}
where in natural units ($\hbar=c=1$),
\begin{equation}
\label{def:gamma}
\gamma =\gamma_0 \, M_{\text{Planck}}^{-2} = \gamma_0 \, l_{\text{Planck}}^2,
\end{equation}
is the Lorentz invariant scale, and $\gamma_0$ is a numerical parameter used to fix the scale . Experimental data and QG theories suggest that the intermediate scale $\gamma$ should be found somewhere between the electroweak length scale $\sim 10^{17}l_{\text{Planck}}$ and the Planck length scale.  

In a recent study~\cite{Nenmeli:2021orl}, we employed the GUP model with maximum momentum uncertainty in Eq. \eqref{def:GUPcomm} to establish a relationship between the GUP-modified dynamics of a massless spin-2 field and Stelle gravity with mass degeneracy~\cite{Stelle1978-du,Stelle1977-rd,Noakes1983-fo}. We also showed that the mass-degenerate Stelle gravity leads to inflation with a natural exit and can be \emph{mapped} to Starobinsky gravity in the sense that the dynamics predicted by both models are the same. Using this, we obtained some of the best-known bounds on GUP parameters~\cite{Nenmeli:2021orl}.

Recently, it was shown that the massive spin-2 modes in the Stelle gravity carry more energy than the scalar modes in Starobinsky model~\cite{2022-Chowdhury.etal}.
It is also known that the Starobinsky model is in perfect agreement with the Planck observations as it predicts a low value of the scalar-to-tensor ratio ($r$)~\cite{Planck:2015sxf,Planck:2018jri,Planck:2018nkj,2021-BICEP3-PLANCK}. 
This leads to the following questions:
Are there signatures distinguishing
mass-degenerate Stelle gravity and Starobinsky model? What are the observable signatures of the GUP modified gravity? In this work, we 
address these questions by investigating the evolution of the scalar and tensor perturbations in these two scenarios.

In order for the inflationary model of the mass-degenerate Stelle gravity to be successful, it must lead to suitable primordial density perturbations that are consistent with the CMB observations~\cite{WMAP:2003elm,Planck:2018nkj}. In this work, we evaluate the spectral tilt of the scalar perturbations $n_{\mathcal{R}}$, and the scalar to tensor ratio $r$. By comparing the primordial scalar power spectrum with the PLANCK~\cite{Planck:2018jri,Planck:2018nkj}, we obtain the bounds on the GUP parameter. Interestingly, the values for $\gamma_0$ we obtain here are consistent with the bounds obtained in Refs.~\cite{Nenmeli:2021orl,Das:2021nbq,Das:2021lrb} and bounds from quantum mechanical considerations~\cite{Todorinov2018-xi,Bosso:2020aqm,Bosso:2020ztk}.

Using the bounds of $\gamma_0$, we evaluate the power spectrum of PGWs generated during inflation. We show that these can be observed in the upcoming gravitational wave detectors, such as ET~\cite{Hild:2010id} and CE~ \cite{Evans:2021gyd,Srivastava:2022slt}. 
Additionally, such detection can provide a direct constrain on the number of e-foldings of inflation. \\[1pt]

\noindent \underline{\sl From Stelle to $f(R)$ + Weyl-squared:}
%
In \cite{Nenmeli:2021orl} we show that the gravitational action derived via Ostrogradsky method from the GUP modified Equations of Motion of a spin-2 field\footnote{Derived from the irreducible representations of a GUP modified Poincar\'e group the EoMs read $\Box h_{\mu\nu}-2\gamma\Box^{2}h_{\mu\nu}=0$.} reads as follows: 
\begin{equation}\label{eq:StelleAction}
    S=\frac{1}{2\kappa}\int \, d^4x \, \sqrt{-g} \left[ 
R  - 2\gamma \, R^{\mu\nu}R_{\mu\nu} + \gamma \, R^{2} 
\right]\, ,
\end{equation}
where $\kappa = 8 \pi G, \gamma$ is defined in Eq. \eqref{def:gamma}, the full procedure is outlined in the Appendix. The above action is identical to Stelle gravity  \cite{Stelle1978-du,Stelle1977-rd,Noakes1983-fo}. Interestingly, the masses of the two additional Yukawa Bosons  \emph{coincide} $(m_{\text{Yukawa}}=1/\sqrt{2 \gamma})$~\cite{Nenmeli:2021orl}. 
The new quadratic terms in the action can be written as the following linear combination 
\begin{equation}
    - 2\gamma \, R^{\mu\nu}R_{\mu\nu} + \gamma \, R^{2} =-\gamma\left[-G^2+C^2 - R^2/{3}\right]\,,
\end{equation}
where $C^2 =  C_{\alpha\beta\rho\sigma} C^{\alpha\beta\rho\sigma} \equiv \mathcal{L}_{\text{Weyl}}$ is the Weyl tensor square~\footnote{We use the following definition of the Weyl tensor in 
4-D:
$
C_{\alpha\beta\rho\sigma}=R_{\alpha\beta\rho\sigma}-\left(g_{\alpha[\rho}R_{\sigma]\beta}-g_{\beta[\rho}R_{\sigma]\alpha}\right)+{\frac {1}{3}}R~g_{\alpha[\rho}g_{\sigma]\beta}\,
$ \cite{hawkinglarge,Wald:1984rg}.}.  $G^2$ is the Gauss-Bonnet invariant, 
%
%
and $R^2$ is the Ricci scalar squared.  Since the Gauss-Bonnet invariant is a boundary term in 
$(3+1)-$dimensions, 
it does not contribute to the dynamics~\cite{Lovelock:1971yv}. Thus, action \eqref{eq:StelleAction} is: 
\begin{flalign}
\label{eq:F(R)} 
S_f=\frac{1}{2\kappa}\int d^4 x \sqrt{-g}\left[f(R)-\gamma\mathcal{L}_{\text{Weyl}}\right];~
f(R)=R+ \frac{{\gamma} R^2}{3} 
\end{flalign}
Using the fact that $\mathcal{L}_{\text{Weyl}}$ is invariant under conformal transformations, we do a series of transformations~\cite{Sotiriou:2008rp}. 
First, we define the inflaton field and its potential as
\begin{equation}
    \phi\equiv\frac{df}{dR}~~ , ~~~V(\phi)\equiv\frac{1}{2\kappa}\left(R\frac{df}{dR}-f\right)\,,
\end{equation}
Next, we perform the following transformation~\cite{Sotiriou:2008rp}: 
%
$\tilde{g}_{\mu\nu}= 
g_{\mu\nu} ({df}/{dR}) 
=\phi g_{\mu\nu}$ ,
%
under which the potential transforms as
\begin{equation}
U\equiv\frac{1}{2\kappa \phi^2}\left(R\frac{df}{dR}-f\right)=\frac{1}{2\kappa}\frac{V}{\phi^2}\,,
\end{equation}
where $f(R)$ is given in Eq. \eqref{eq:F(R)}. Lastly, we redefine the scalar field $\phi$ as:
%
$    \chi\equiv \ln{\left[\phi\right]}/\sqrt{3\kappa}$,
    %
for which the potential $U(\chi)$ takes the form
\begin{equation}
    U(\chi)=\frac{3}{4}\frac{M_{\text{Planck}}^2}{8\gamma}\left(1-e^{\sqrt{2/3}\chi/M_{\text{Planck}}}\right)^2\,.
\end{equation}

Thus, the action \eqref{eq:StelleAction} is transformed to:
\begin{equation}\label{eq:f(r)+Weyl}
\tilde{S}=\int \!\! d^4 x \sqrt{-\tilde{g}} \left[\frac{\tilde{R}}{2\kappa}-\frac{\tilde{g}^{\mu\nu}}{2}\partial_\mu\chi\partial_\nu\chi-U(\chi)-\frac{\gamma}{2\kappa}\mathcal{L}_{\text{Weyl}}\right],
\end{equation}
in which the scalar field has been separated from the curvature term.
The equations of motion reduce to
\begin{eqnarray}
\label{eq:StelleEOM01}
& & \tilde{G}_{\mu\nu} - 2\gamma \tilde{B}_{\mu\nu} 
= \kappa^2  T^{\chi}_{\mu\nu},
\end{eqnarray}
where $G_{\mu\nu}$ is the Einstein tensor, Bach tensor ($B_{\mu\nu}$) is 
%
\begin{equation}\label{eq:BachTensorDef}
   \tilde{ B}_{\mu\nu}\equiv\tilde{\nabla}^\rho\tilde{\nabla}^\sigma \tilde{C}_{\mu\rho\nu\sigma}-\frac{1}{2}\tilde{R}^{\rho\sigma}\tilde{C}_{\mu\rho\nu\sigma},
\end{equation}
and $T_{\mu\nu}^{\chi}$ is the stress-energy tensor of the $\chi$ field which has the same form as a scalar field stress-energy tensor
\begin{equation}
T_{\mu\nu}^{\chi}\equiv\partial_\mu\chi\partial_\nu\chi-\delta_{\mu\nu}\left(\frac{1}{2}\tilde{g}^{\rho\sigma}\partial_\rho\chi\partial_\sigma\chi+U(\chi)\right)\,.
\end{equation}
From Eq. \eqref{eq:StelleEOM01} we see that the scalar
field ($\chi$) and the geometrical part can be quantized independently. 

As mentioned earlier, in Ref.~\cite{Nenmeli:2021orl}, we analytically showed that Stelle gravity, when applied to a homogeneous, isotropic background, leads to inflation with exit. In this work, to connect to observables, we consider the following perturbed FRW line-element~\cite{1984-Kodama.Sasaki-PTPS,1992-Mukhanov.etal-PRep}:
\begin{eqnarray}
\label{eq:per-frw}
{\rm d}s^2 &=& a^2(\eta )\left[- (1+2\varphi){\rm d}\eta ^2 + 2 \partial_i B {\rm d}x^i
{\rm d}\eta \right. \\
& & + \left. \left[(1+2 \psi )\delta _{ij}+2 \partial_i \partial_j E + h_{ij} \right]
{\rm d}x^i{\rm d}x^j \right]\ , \nonumber
\end{eqnarray} 
where the functions $\varphi$, $B$, $\psi$ and $E$ represent the scalar
sector whereas the tensor $h_{ij}$, satisfying $h_i^i = \partial^{i} h_{ij}
= 0$, represent gravitational waves. $a(\eta)$ is the expansion factor in conformal time $\eta$. Note that all these first-order
perturbations are functions of $(\eta, {\bf x})$. For convenience, we
do not write the dependence explicitly.
Interestingly, the GUP modified action \eqref{eq:StelleAction}, like Einstein gravity, leads to scalar and tensor fluctuations~\cite{Anselmi:2020lpp}. In other words, GUP modified action does not contain any growing vector
fluctuations, or additional scalar and tensor fluctuations.

Substituting the above line-element \eqref{eq:per-frw} in Eq. \eqref{eq:StelleEOM01}, scalar perturbation equation (in Fourier 3-space) is~\cite{Deruelle:2010kf,Deruelle:2012xv}: 
{\small
\begin{equation}
\gamma\,
\left[W^{(4)} - \frac{\ddot H}{\dot H} W^{(3)}\right] 
+ C_2\,\ddot W
+ C_1\,\dot W
+ C_0\,W
= 0\,,
\label{eq:eom-scalar2}
\end{equation}
}
where $W=\psi-\varphi$, $H$ is the background Hubble parameter, and $\{C_0,C_1,C_2\}$ are coefficients determined by the wavenumber of the scalar perturbations $k$, the QG scale $\gamma$, the scale factor $a$, the Hubble parameter $H$, and their derivatives. The EoMs of the tensor perturbations are given by:
\begin{equation}
\Box \bar h_{ij} - 2 \mathcal H\,\bar h_{ij}'
- \frac{\gamma}{a^2} \Box^2 \bar h_{ij}=0\, . 
\label{eq:eoms-tensor}
\end{equation}
In the rest of this work, we use two distinct methods to connect to observables. First, we use the Fakeon procedure~\cite{Anselmi:2020lpp} to constraint the GUP parameter from the CMB observations~\cite{Planck:2018nkj}. Later, using the constraint on $\gamma_0$ from CMB observations, we derive the power spectrum of PGWs~\cite{Deruelle:2010kf,Deruelle:2012xv}, and compare with the design sensitivity of ET~\cite{Hild:2010id} and CE~\cite{Evans:2021gyd,Srivastava:2022slt}.

\noindent \underline{\sl Observables:}
Recently, imposing 
constraints of locality, unitarity, and renormalizability for quantum gravity, and using the fakeon procedure~\cite{Anselmi:2017ygm}, 
it was shown that quantum gravity could contain only two more independent parameters 
than Einstein gravity. This was identified as the inflaton $\chi$ and the spin-2 fakeon $\chi_{\mu\nu}$~\cite{ Anselmi:2020lpp}. Interestingly, the extra degrees of freedom are turned into fake ones and projected away. Note that the Stelle action and the $f(R)+C^2$ action are renormalizable, although they contain Fadeev-Popov (non-malicious) ghosts.

Repeating the analysis~\cite{Sasaki2003-of,Anselmi:2020lpp},  yields the following expressions for the power spectra of the scalar and tensor perturbations, respectively:
\begin{eqnarray}
\mathcal{P}(k)_{\mathcal{R}}= A_{\mathcal{R}}\left(\frac{k}{k^*}\right)^{n_{\mathcal{R}}-1} &;&
\mathcal{P}(k)_{T} = A_{T}\left(\frac{k}{k^*}\right)^{n_T}\,,
\end{eqnarray}
where $k_*$ is the wave-number of the fluctuations at the horizon crossing. ${P}(k)_{\mathcal{R}}$, $A_{\mathcal{R}}$ and $n_{\mathcal{R}}$ are the power spectrum, the amplitude, and the spectral tilt of the scalar perturbations respectively. Analogously, ${P}(k)_{T}$, $A_{T}$, and $n_T$ are the power spectrum, the amplitude, and the spectral tilt of the tensor perturbations. We will obtain these quantities during slow-roll inflation~\cite{1992-Mukhanov.etal-PRep,1995-Lidsey.etal-RMP}. 

\begin{figure}[ht]
\centering
\includegraphics[height=0.25\textwidth]{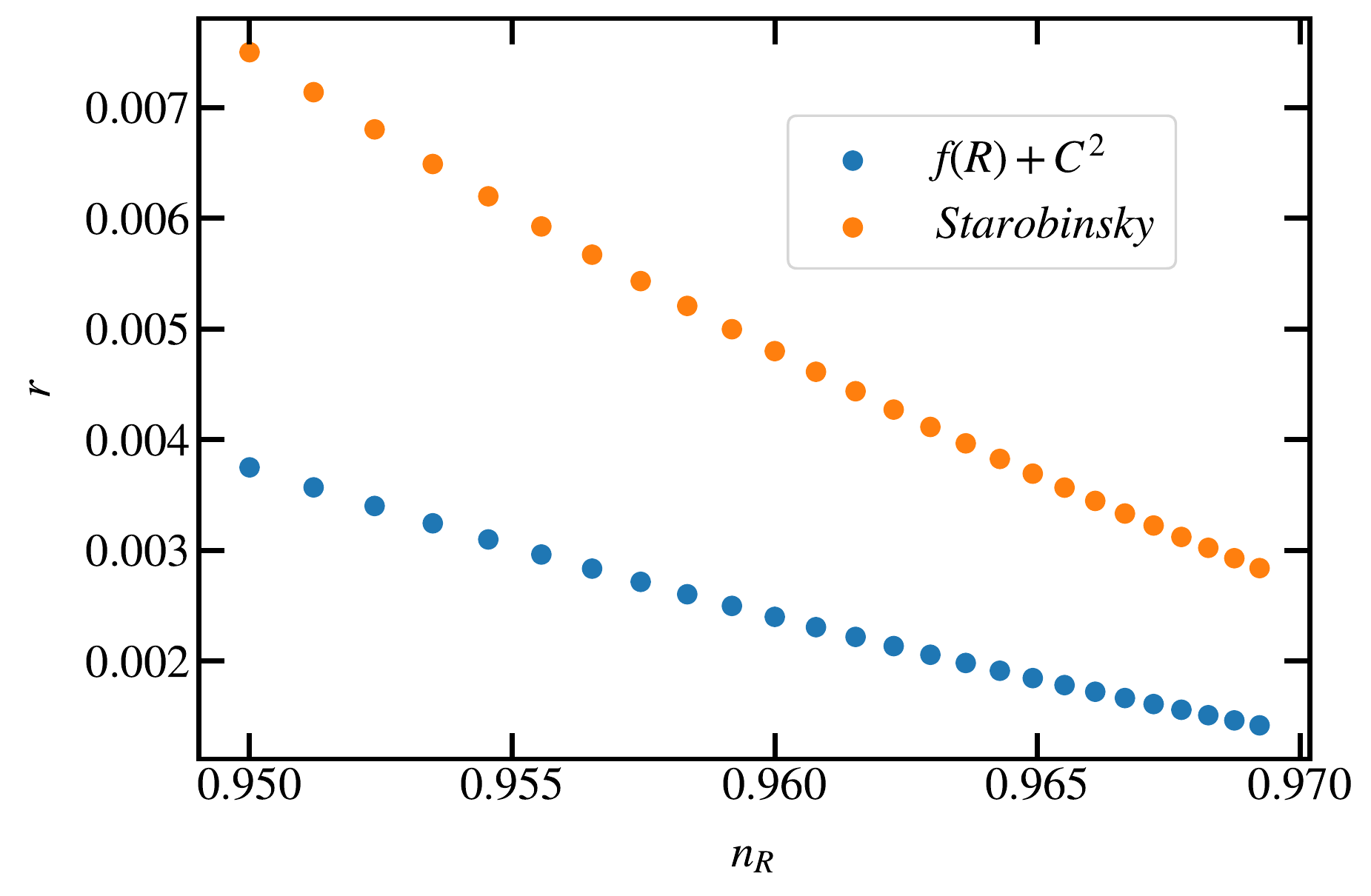}
  \caption{The graph represents the spectral tilt of the scalar perturbations $n_{\mathcal{R}}$ and the  tensor to scalar ratio $r= A_{T}/A_{\mathcal{R}}$ from $N=40$  to $N=70$ number of e-folds. One can see the QG suppression of the tensor perturbations.}\label{fig:one}
    \end{figure}

During the slow-roll inflation, the number of e-folds of inflation $N$ can be rewritten in terms of the slow-roll parameter ($\varepsilon$):
\begin{equation}
N=\int_{\varepsilon }^{1}\frac{H(t (\varepsilon  ))}{\dot{%
\varepsilon}(t (\varepsilon  ))}\mathrm{d}\varepsilon
\simeq \frac{1}{2\varepsilon }-\frac{1}{12}\ln
\varepsilon +\text{ }\mathcal{O}(\varepsilon ^{0}).  \label{N}
\end{equation}%
Thus, in the leading order, $\varepsilon$ and $N$ are inversely proportional, i. e.,
%
$\varepsilon\approx {1}/{2 N}$.
%
Up to the leading order, the amplitudes and the spectral tilts are given by
%
\begin{eqnarray}
& & A_{\mathcal{R}} =\frac{Gm_{\phi }^{2}}{12\pi \varepsilon^2 }\left( 1-\sqrt{%
3\varepsilon }-(n_{\mathcal{R}}-1)(2-\gamma _{E}-\ln 2)\right) \nonumber \\
\label{eq:observables}
&& n_{\mathcal{R}}-1 =-4\sqrt{\frac{\varepsilon }{3}}, 
\qquad n_{T} = - \frac{4}{3} \varepsilon \\
& & A_{T} = \frac{8Gm_{\phi }^{2}}{3 \pi} \left[1 - \frac{2}{\sqrt{3}} \sqrt{\varepsilon} - \frac{39}{54} \varepsilon -n_{T}(2-\gamma
_{E}-\ln 2)\right] \nonumber \\
& &
%
%
r=\frac{A_{T}}{A_{\mathcal{R}}}=
\frac{32 \varepsilon}{3} 
\left[1 + \sqrt{\frac{\varepsilon}{3}} + [n_{\mathcal{R}}-1-n_{T}][2-\gamma _{E}-\ln
2]\right] \nonumber
%
\end{eqnarray}
%
where $m_{\phi}=1/\sqrt{2\gamma}$ is the mass of the spin-0 mode,  
and $\gamma_E$  is the  Euler-Mascheroni constant. Since Bach tensor is conformally invariant, it vanishes 
in the conformal flat FRW geometry. 
To the leading order, the scalar (${P}(k)_{\mathcal{R}}$) and tensor (${P}(k)_{T}$) power-spectrum  are: 
%
\begin{equation}
\label{eq:TheoreticalAs}
\!\!\!\!\! A_{\mathcal{R}}
=\frac{N^2}{18\pi\gamma_0 };~
A_{T} 
=\frac{1}{\pi\gamma_0 }  \, .
\end{equation}
This is the first key result regarding which we would like to discuss the following points: First, 
the spectral tilt $n_{\mathcal{R}}$ and $n_{T}$, and the scalar-to-tensor ratio $r$, depend on $\gamma_0$. However, it does not contain additional deviations from the 
Bach tensor. Second, from the above expressions on $A_{\mathcal{R}}$ and $A_{\mathcal{T}}$, we can derive bounds on the magnitude of $\gamma_0$. We obtain this by comparing Eq. \eqref{eq:TheoreticalAs} with PLANCK observations~\cite{Planck:2015sxf,Bonga:2015xna,Planck:2018nkj}. Additionally, we compare the QG modified observables with Starobinsky model~\cite{1980-Starobinsky-PLB}. This is because the Starobinsky model perfectly agrees with the Planck observations as it predicts a low value of the scalar-to-tensor ratio and mass-degenerate Stelle gravity leads to inflation with a natural exit and can be \emph{mapped} to Starobinsky gravity.
Third, Fig.~\ref{fig:one} contains the comparison of the spectral tilt of the scalar modes $n_{\mathcal{R}}$ and the scalar-tensor ratio $r$ in the Starobinsky and $f(R)+C^2$ gravity model. While the spectral tilt of the scalar power spectrum is the same in both cases, the introduction of $R_{\mu\nu}R^{\mu\nu}$ decreases the ratio between the scalar contribution and the tensor modes. This is because Stelle gravity contains extra massive spin-2 modes and they carry more energy compared to the scalar modes. This is consistent with the recent analysis where it is shown that QG effects suppress the tensor modes~\cite{2022-Chowdhury.etal}.
Finally, the bounds on the GUP parameter $\gamma_0$ can be obtained from the PLANCK observations~\cite{Planck:2015sxf,Bonga:2015xna,Planck:2018nkj}:
\begin{equation}\label{eq:ExperimentalAs}
    A_{\mathcal{R}} = 2.474 \pm 0.116 \times 10^{-9}\,.
\end{equation}
Comparing the theoretical results of Eq. \eqref{eq:TheoreticalAs} with the above expression leads to the following value for $\gamma_0$ for two different e-folds ($N=40$ and $N=60$) of inflation:
\begin{equation}
\label{eq:gamma0values}
\gamma_{0}^{N=40} \approx 3.430\times 10^{10} \, , ~~\gamma_{0}^{N=60}\approx 7.719\times 10^{10}.
\end{equation}
Thus, the values for the intermediate scale are consistent with the bounds obtained in Refs.~\cite{Nenmeli:2021orl,Todorinov2018-xi,Bosso:2020aqm,Bosso:2020ztk,Das:2021lrb}.

\begin{figure}
\centering
         \centering       \includegraphics[height=0.25\textwidth]{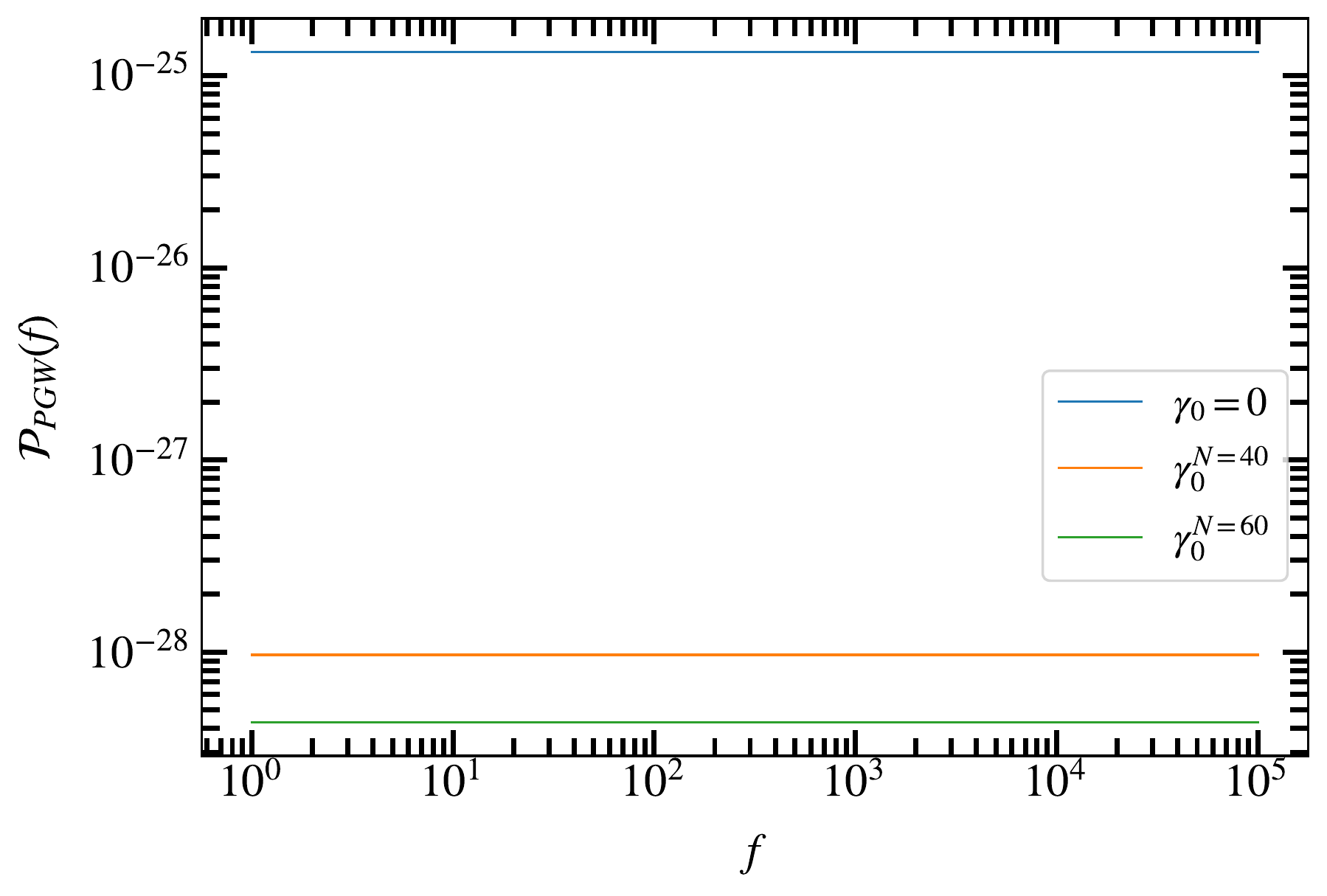}
         \includegraphics[height=0.25\textwidth]{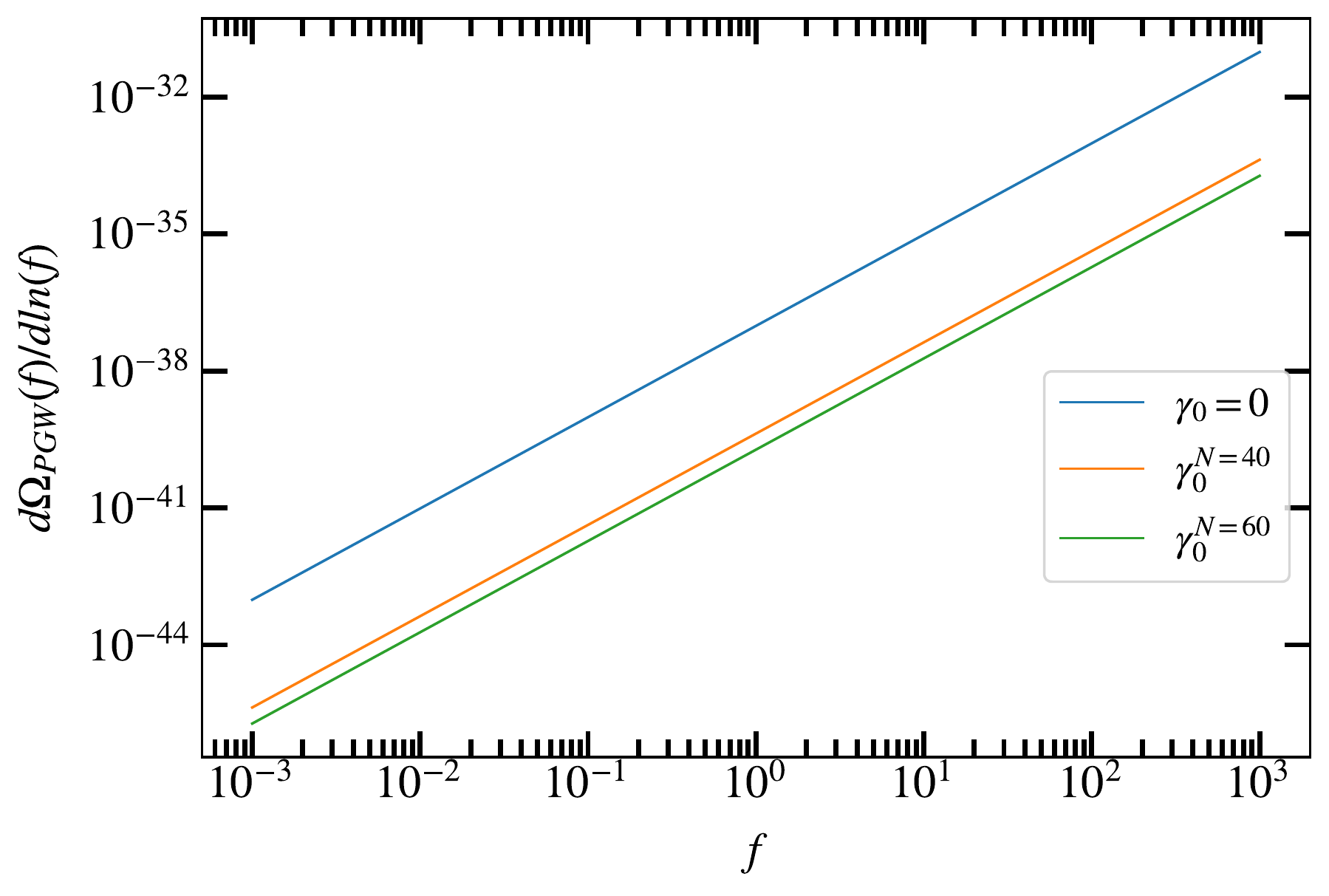}
         \caption{The top graph shows a logarithmic plot of the power spectrum $\mathcal P_{\rm PGW}(f)$ as a function of the frequency $f$ and the GUP parameter $\gamma_0$. The bottom graph shows the differential energy density of PGWs as a function of frequency.}
         \label{fig:two}
\end{figure}

\noindent \underline{\sl Power spectrum of PGWs:} 
%
The next generation of ground-based GW detectors~\cite{Hild:2010id,Evans:2021gyd,Srivastava:2022slt} and LISA~\cite{2017-LISA} 
are expected to deliver data that will help to answer some of the deep questions in  fundamental physics, astrophysics, and cosmology~\cite{2019-Barack.etal-CQG}. Modified theories of gravity have shown an impact on the PGWs power spectrum \cite{Oikonomou:2022pdf,Odintsov:2022cbm}. Here, we show further that the quantum gravitational signatures in PGWs can potentially be observed in the next-generation gravitational wave detectors.  
Rewriting Eq.~\eqref{eq:eoms-tensor} as two second-order differential equations, 
in the Fourier domain, we have~\cite{Deruelle:2010kf,Deruelle:2012xv}:
\begin{equation}
\begin{aligned}
\frac{d^2\mu^{(1)}_k}{dz^2}
+ \left(1-\frac{2}{z^2}\right)\,\mu^{(1)}_k
= 0\,, \\
\frac{d^2\mu^{(2)}_k}{dz^2}
+ \left(1+\frac{1}{\gamma\,H^2\,z^2}\right)\,\mu^{(2)}_k
= 0\,,
\end{aligned}
\label{eq:eomWbis}
\end{equation}
where $\mu^{(1)}$ ($\mu^{(2)}$) corresponds to the standard gravitational wave (massive spin-2 mode from the modified theory), and $z\equiv -k\eta$.
For pure de Sitter, the power spectrum can be evaluated exactly~\cite{2004-Shanki.Sriram-PRD,2006-Sriram.Shanki-JHEP}  and is given by:
\begin{align}
\mathcal P_{\rm PGW}(k)
&= \mathcal P_0(\gamma_{0}, H) 
  \left[
   1 + \left[\frac{k}{k_*}\right]^2 \right.\\
&\left. 
  - \frac{\pi}{2}\,\left[\frac{k}{k_*}\right]^3 \!\!
     |\mathrm e^{\mathrm i \pi\,\nu/2}\,H^{(1)}_{\nu}\left[\frac{k}{k_*}\right]|^2
  \right],\label{eq:PowerSpectrumQG}
\end{align}
where $\mathcal P_0(\gamma_{0}, H) = (2 \kappa\,H^2/\pi^2)(1 + 2 \gamma\,H^2)$, $k_{*}=aH$ is the wavenumber at horizon crossing, $H^{(1)}_{\nu}$ is the Henkel function of the first kind, $H$ is the Hubble scale during inflation and $\nu=\frac{1}{4}\sqrt{1-\frac{4}{\gamma H^2}}$. The power spectrum given in Eq. \eqref{eq:PowerSpectrumQG} is valid for $\gamma_0>0$. For $\gamma_0 = 0$, the unmodified power spectrum for the PGWs is:
\begin{equation}
  \mathcal P_{\rm PGW}(k)=\mathcal P_0(\gamma_{0}, H) \left(\frac{k}{k_*}\right)^{n_T}\,.
\end{equation}
Evaluating the power spectrum at the horizon crossing (i.e., $\eta\rightarrow 0$), for three values of $\gamma_0$ --- 
$\{\gamma_0 = 0, \gamma_{0}^{N=40}$ and $\gamma_{0}^{N=60}\}$~\eqref{eq:gamma0values} --- we obtain the following values for the normalization constant 
$\mathcal P_0(\gamma_{0}, H)$: 
\begin{subequations}
\begin{align}
\mathcal P_0(\gamma_{0}=0, H = 10^{15} {\rm GeV})&=1.329\times10^{-25}\\
  \mathcal P_0(\gamma_{0,N=40}, H = 10^{15} {\rm GeV}) &=4.841\times 10^{-29}\\
    \mathcal P_0(\gamma_{0,N=60}, H = 10^{15} {\rm GeV}) &=2.152\times 10^{-29}\, ,
\end{align}
\end{subequations}
%
%
Fig. \ref{fig:two} contains the plot of the power spectrum of PGWs (for three different values of $\gamma_0$) as a function of frequency. From the figure, we infer that the power spectrum is  
almost constant for a large range of frequencies and is consistent with standard inflationary scenario~\cite{1997-Turner-PRD}. Also, the degenerate Stelle gravity model suppresses the 
amplitude of PGWs. This implies that the longer the
inflation, the larger the suppression of the 
amplitude of the PGWs. This is because the massive spin-2 modes carry less energy than the scalar
modes~\cite{2022-Chowdhury.etal}. 
Further, to quantify the detectability of PGWs in the upcoming GW detectors, we evaluate the energy density of PGWs~\cite{Smith:2005mm}: 
\begin{equation}\label{eq:GWEnergyDensity}
  \Omega_{\text{GW}}(k) = \mathcal P_{\rm PGW}(k) \, (k^2/{(12 H_0^{2})})\, .
\end{equation}
where we have set the scale factor at the present time to unity,
 and $H_0=100 \,h\, \text{km}\,\text{s}^{-1}\,\text{Mpc}^{-1}$. 
 Figure \ref{fig:two} contains the plot of 
differential energy density is ${d (\Omega_{\text{GW}}(k))}/d\ln(k)$ as a function of $f$. 
From the figure, we infer that the differential energy density is 8-orders of magnitude larger in $kHz$ compared to $1~Hz$ range for all three values of $\gamma_0$. To confirm this, we now
compare the projected characteristic strain for PGWs and the characteristic strain of the detectors. The characteristic strain of the PGWs is given by~\cite{Moore:2014lga}:
\begin{equation}\label{eq:CharStrain}
    h_{PGW}= \int \mathcal P_{\rm PGW}(f) \, df\,.
\end{equation}
where $\mathcal P_{\rm PGW}(f)$ is given by Eq. \eqref{eq:PowerSpectrumQG}. Since 
$\mathcal P_{\rm PGW}(f)$ is approximately constant with frequency (cf. 
 Figure \ref{fig:two}), we approximate $\mathcal P_{\rm PGW}(f)\approx   \mathcal P_0(H, \gamma_0)$. Thus, we get: 
\begin{equation}\label{eq:CharStrainApprox}
     h_{PGW}\approx f \, \mathcal P_0(H, \gamma_0)\,.
\end{equation}

%

\begin{figure}[h]       
    \centering
\includegraphics[height=0.3\textwidth]{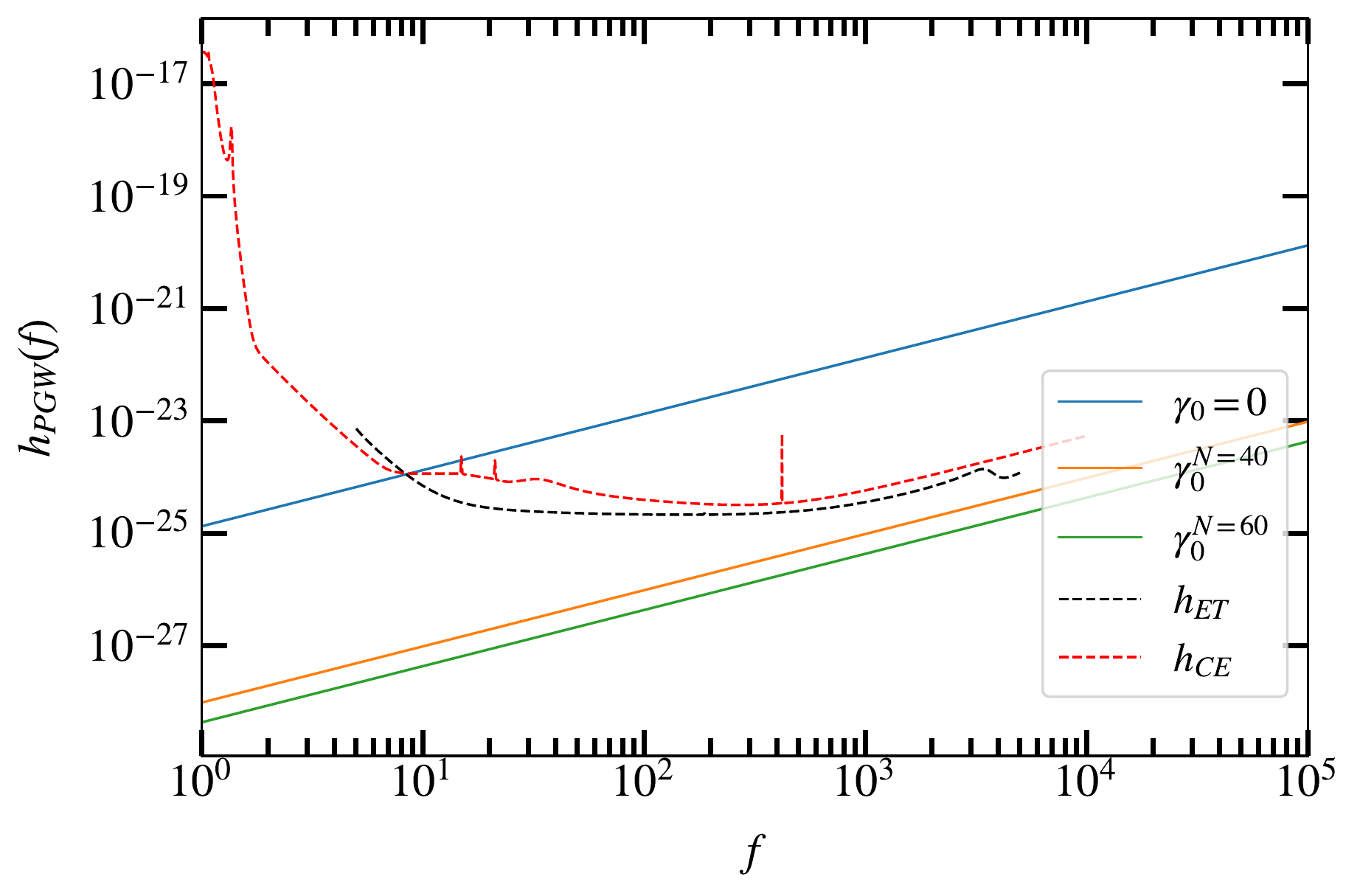}
         \caption{Logarithmic graph of the characteristic strain of the gravitational waves produced during inflation $h_{GW}$ as a function of frequency $f$ and the GUP parameter $\gamma_0$, compared to the characteristic strain of ET $h_{ET}$ \cite{Hild:2010id} and CE $h_{CE}$ \cite{Evans:2021gyd,Srivastava:2022slt}.}
         \label{fig:three}
     \end{figure}
Figure \ref{fig:three} contains the  characteristic strain of PGWs for three different values of $\gamma_0$ (corresponding to different e-foldings of inflation) along with the design sensitivity of ET~ \cite{Hild:2010id} and CE~ \cite{Evans:2021gyd,Srivastava:2022slt}.  From the figure, we infer the following: First, the characteristic strain of PGWs for the standard inflation (blue curve) is well within the observational capability of both ET and CE. Second, for degenerate Stelle gravity, as the number of e-foldings increases, the suppression of the 
amplitude of the PGWs is larger (orange and green curves). Hence, it is not possible to confirm their detection with confidence. However, the non-detectability of the PGWs will directly constrain the value of $\gamma_0$. Thus, ET~\cite{Hild:2010id} and CE~\cite{Evans:2021gyd,Srivastava:2022slt} can put a severe constraints on the value of the GUP parameter. Third, our analysis shows that LISA will not be able to detect PGWs as the characteristic strain is much larger than PGWs in low-frequency ($\sim 1~{\rm Hz}$). Lastly and most importantly, our analysis explicitly shows that we can strongly constrain the number of e-foldings of inflation from ET and CE.

\noindent \underline{\sl Conclusions}
In conclusion, we demonstrated that our model is described by a linear combination of $f(R)$ gravity, particularly the Starobinsky model and Weyl-squared gravity. This allowed us to use the fakeon procedure~\cite{Anselmi:2020lpp} to obtain the scalar spectral tilt $n_{\mathcal{R}}$ and the tensor-scalar ratio $r$. Although our model and Starobinsky model are identical in the background FRW universe, they lead to different observable consequences in the linear order in perturbations. More specifically, we showed that the introduction of $R_{\mu\nu}R^{\mu\nu}$ decreases the ratio between the scalar contribution and the tensor modes. This is consistent with the recent  analysis~\cite{2022-Chowdhury.etal}. Using PLANCK observations, we determined the bounds on the GUP parameter $\gamma_0$. These bounds are consistent with bounds derived from other analysis~\cite{Nenmeli:2021orl,Todorinov2018-xi,Bosso:2020aqm,Bosso:2020ztk,Das:2021nbq,Das:2021lrb}.

Later, using the constraint on $\gamma_0$ from the CMB observations, we explicitly showed that PWGs carry quantum gravitational signatures in their power spectrum and energy density. Furthermore, from Figure \ref{fig:three} one can conclude that these quantum gravitational effects will potentially be observable in the next-generation ground-based gravitational wave detectors such as ET and CE. Additionally, depending on the energy density of detected PGWs, one can strongly constrain the number of e-foldings of inflation using these detectors.

After the historical detection of GWs in the frequency range $100 - 10^4~{\rm Hz}$, there is a surge in activity for the possibility of detection of GWs in the MHz-GHz frequency range~\cite{2020-Aggarwal.etal-arXiv}. As mentioned earlier, for a large frequency range, $\mathcal P_{\rm PGW}$ is approximately constant. This means that high-frequency GW detectors with sensitivity around $10^{-20}$ can potentially detect PGWs. Such detectors will reveal a disparate view of the early universe compared to their electromagnetic counterparts. 

\noindent {\bf Acknowledgments:} 
The authors thank Avijit Chowdhury, Ashu Kushwaha, Abhishek Naskar and Vijay Nenmeli for their comments on the earlier draft. This work was supported by the 
Natural Sciences and Engineering Research Council of Canada. 
SS is supported by SERB-MATRICS grant.

\section*{Appendix} 
\label{Appendix}

From the Relativistic GUP (RGUP) relation \eqref{def:GUPcomm}, it is clear that the position and momentum operators are no longer canonically conjugate. The introduction of a "canonical" 4-momentum $p_{0}^{\mu}$ satisfying $[x^{\mu},p_{0}^{\nu}]=i\eta^{\mu\nu}$ simplifies calculations considerably. The "physical" 4-momentum $p^{\mu}$ can be expressed in terms of $p_{0}^{\mu}$ as:
\begin{equation}
    \label{pFromp0}
      p^{\mu}=p_{0}^{\mu}(1-\gamma p_{0\nu}p_{0}^{\nu}).
\end{equation}
In Ref.~\cite{Nenmeli:2021orl}, we used the  field theoretic approach to obtain a one-to-one correspondence between RGUP modified Spin-2 field theory and modified gravity: 
\begin{enumerate}
\item The Lagrangian for a \textit{free} spin-2 field must be bilinear in the field and its derivatives, $h_{\mu\nu}$. Relevant bilinears are readily enumerated.
\item Due to the long range of gravitational force, the mediating gauge bosons must be massless. Thus, we can assume that the Lagrangian consists solely of field derivatives and has no mass term.
\item We can create a \textit{minimal} list by identifying pairs of bilinears that differ by surface terms. Thus, we can fix the action up to coefficients that are undetermined: 
 \begin{equation}
    \label{Spin2Lag1} \mathcal{L}=ah_{\mu\nu,\sigma}h^{\mu\nu,\sigma}+bh_{\mu\nu}\,^{,\nu}h^{\mu\sigma}\,_{,\sigma}+ch_{\mu\nu}\,^{,\nu}h^{\sigma}\,_{\sigma,\mu}+dh_{\mu}\,^{\mu}\,_{,\sigma}h^{\nu}\,_{\nu}\,^{,\sigma} \nonumber
\end{equation} 
\item An interaction term of the form $\lambda h_{\mu\nu}T^{\mu\nu}$ takes into account the matter-gravity interactions. By adding this term to the preceding action and enforcing energy-momentum conservation (i.e. $\partial_{\mu}T^{\mu\nu}=0$), 
we obtain the following result:
\begin{equation}
    \label{Spin2Lag2}
    (2a+b)\Box h_{\mu\nu}\,^{\nu}+(b+c)h_{\nu\sigma}^{,\nu\sigma}\,_{,\mu}+(c+2d)\Box h_{\nu}\,^{\nu}\,_{,\mu}=0
    \end{equation}
We can use the above expression to fix the coefficients. 
\end{enumerate}
The Lagrangian corresponding to the above EOM is 
\begin{equation}
\label{Spin2Lag3}
    \mathcal{L}_{free}=\frac{1}{2}h_{\mu\nu,\sigma}h^{\mu\nu,\sigma}-h_{\mu\nu}\,^{,\nu}h^{\mu\sigma}\,_{,\sigma}+h_{\mu\nu}\,^{,\nu}h^{\sigma}\,_{\sigma,\mu}-\frac{1}{2}h_{\mu}\,^{\mu}\,_{,\sigma}h^{\nu}\,_{\nu}\,^{,\sigma}
\end{equation}
The RGUP modified EOM are obtained from the position space representation of \eqref{pFromp0} (i.e. $\partial_{\mu}\rightarrow\partial_{\mu}(1-\gamma\Box)$). We have
\begin{equation}
\label{Spin2EOMMod1}
G^{L}_{\mu\nu}+\gamma\mathcal{G}_{\mu\nu}=0,
\end{equation}
        where 
\begin{eqnarray}
\label{Spin2EOM}
& & \!\!\!\!\!\!\!\!\!\!\!\!\!\!\!\!\!\!\!\!\!\!\!\!\!  
G^{L}_{\mu\nu} =  \Box h_{\mu\nu}-h_{\mu\sigma,\nu}^{\sigma}-h_{\nu\sigma,\mu}^{\sigma}+h_{\sigma,\mu\nu}\,^{\sigma}+\eta_{\mu\nu}h_{\sigma\rho}^{,\sigma\rho}-\eta_{\mu\nu}\Box h_{\sigma}^{\sigma}, \\
\label{Spin2EOMMod2}
 & & \!\!\!\!\!\!\!\!\!\!\!\!\!\!\!\!\!\!\!\!\!\!\!\!\!
\mathcal{G}_{\mu\nu} = \Box^{2}h_{\mu\nu}-\Box h_{\mu\sigma,\nu}^{,\sigma}-h_{\nu\sigma\mu}^{,\sigma}.    +\Box h_{\sigma,\mu\nu}^{,\sigma} +\eta_{\mu\nu}h_{\sigma\rho}^{,\sigma\rho}
-\eta_{\mu\nu}\Box^{2}h_{\sigma}^{\sigma}.~~
\end{eqnarray}
The above linearized equation of motion  
is identical to the equations of motion 
obtained by linearlized Stelle action~\cite{Stelle1977-rd,Nenmeli:2021orl}: 
\begin{equation*}
S=\int \, d^4x \, \sqrt{-g} \left[ 
\frac{1}{2\kappa^{2}}R  - \alpha \, R^{\mu\nu}R_{\mu\nu} + \beta \, R^{2} 
\right] \, .
 \end{equation*}
We showed that the linearized Stelle gravity equations of motion match \eqref{Spin2EOMMod1} perfectly when $\alpha=2\beta=\gamma/\kappa^{2}$. In other words, we showed that $\alpha=2\beta$ Stelle gravity is the \textit{minimally modified, metric-only} theory of gravity which models the effects of maximal momentum. Hence, the model we have used is a one-parameter model, and this equality is then translated into the $f(R)+C^2$ model that we used for the modeling of our inflation parameters.

\bibliography{Ref}
\end{document}


\section{$f(R)$ gravity: Jordan and Einstein frame}
\label{sec:f(R)gravity}

In $f(R)$ gravity one assumes that the Einstein Hilbert action is a first order approximation of the most general action which is a function of Ricci scalar. In this section, we have followed Ref. \cite{1992-Mukhanov.etal-PRep}. 

Let us consider the following action
\begin{multline}
{\cal S}= \frac{1}{2\kappa} \int{d^4x\sqrt{-g} f(R)} +\int d^4x \mathcal{L}_M(g_{a b},\psi_M)\\\tcr{-\frac{\gamma}{2\kappa}\int dx^4 \sqrt{-g} C_{\alpha\beta\rho\sigma} C^{\alpha\beta\rho\sigma}} 
\label{eq:actionJordanfofR}
\end{multline}
\subsection{Field equations}
Varying the action with respect to metric the field equations are given by
\begin{multline}
\Sigma_{p q}=F(R)R_{p q} - \frac{1}{2}g_{p q}f(R)\\-\nabla_p\nabla_q F(R) 
+ g_{p q} \Box{F(R)}\tcr{- 2\gamma B_{p q}}=\kappa T^{(M)}_{p q} \label{eq:fieldfofR}
\end{multline}
where \tcr{$B_{pq}$ is defined in Eq. \eqref{eq:BachTensorDef}}, $F(R)\equiv \frac{\partial{f(R)}}{\partial R}$ and $T^{(M)}_{p q}$ is the energy-momentum tensor of the matter field defined by
\begin{equation}
T^{(M)}_{p q}=-\frac{2}{\sqrt{-g}}\frac{\delta \mathcal{L}_M}{\delta g^{p q}}
\end{equation}
The continuity equation is given by 
\begin{equation}
\label{eq:chap2-cont}
\nabla_{p}T^{p q (M)}=\nabla_p \Sigma^{p q}=0
\end{equation}
\tcr{One can see from the formula for divergence of the Bach tensor in any dimensions that in our case where $n=4$ the Bach tensor is divergence free which can be seen in \cite{GRAHAM20091956} }
\begin{equation}
   \tcr{ \nabla^j B_{ij} = (n-4)P^{jk} ( \nabla_i P_{jk} - \nabla_j P_{ik} )\,,}
\end{equation}
\tcr{where $n$ is the number of spacetime dimensions, and the Schouten tensor $P_{ab}$ is given by
\begin{equation}
    P_{{ab}}={\frac  {1}{n-2}}\left(R_{{ab}}-{\frac  {R}{2(n-1)}}g_{{ab}}\right)
\end{equation}}
From the trace of the equation~(\ref{eq:fieldfofR}) we get
\begin{equation}
3\Box F(R) + R F(R)-2f(R)=\kappa T\,,
\end{equation}
\tcr{which is exactly what we have if we consider only $f(R)$ gravity. This is due to the fact that the Bach tensor is a tracefree tensor \cite{GOVER2006450}}.

Then it is still possible to write the field equations in the form of 
\begin{equation}
G_{p q}\tcr{- 2\gamma B_{p q}} =T^{(M)}_{p q}+T^{(F)}_{p q}
\end{equation}
where $T^{(F)}_{p q}$ is the effective energy-momentum tensor from the modified gravity. Using the fact that 
$\nabla_{p}G^{p q}=0$, \tcr{$\nabla_{p}B^{p q}=0$} and from continuity equation (\ref{eq:chap2-cont}), we get, $\nabla_p T^{p q (M)}=0$. Hence, 
the continuity equation holds for the effective energy-momentum tensor.

\subsection{Conformal transformation}
The $f(R)$action corresponds to a non-linear function of $R$. Also, the field equations for such an action contain higher derivative terms, which makes the analysis difficult. However, using the conformal transformation,
\begin{equation}
\tilde{g}_{p q}=\Omega^2 g_{p q}
\end{equation}
it is possible to obtain the action in Einstein frame where the field equations are second order.

\tcr{In $f(R)+C^2$ case the field equations does not reduce to second order, however we have simplified the left hand side.} 

Here, $\Omega$ is the 
conformal factor. We use a tilde throughout this section to denote quantities in Einstein frame.

The Ricci scalar in both --- Einstein and Jordan --- frames are related by

\begin{equation}
R=\Omega^2\left(\tilde{R}+6\tilde{\Box}\Omega - 6 \tilde{g}^{p q} \partial_{p}{\Omega}\partial_{q}{\Omega}\right)
\end{equation}
\tcr{The connection between Einstein and Jordan frames for the Weyl part of the action follows from
\begin{subequations}
\begin{align}
\tilde{g}_{p q}&=\Omega^2 g_{p q}\\
    g^{p q}&=\Omega^{-2} \tilde{g}^{p q}\\
    \sqrt{-\tilde{g}}&=\Omega^{n}\sqrt{-g}\,,\\
    {C^a}_{\,bcd}&={\tilde{C}^a}_{bcd}\\
    C_{abcd}C^{abcd}&=g_{am}{C^m}_{bcd}\,g^{ak}g^{bl}g^{cp}g^{dq}g_{kn}\,{C^n}_{lpq}\\
    &=\Omega^4\tilde{g}_{am}{\tilde{C}^m}_{bcd}\,\tilde{g}^{ak}\tilde{g}^{bl}\tilde{g}^{cp}\tilde{g}^{dq}\tilde{g}_{kn}\,{\tilde{C}^{n}}_{lpq}
\end{align}
\end{subequations}
where again $n$ is the number of dimensions. We can easily conclude that 
\begin{equation}
    \sqrt{-g} C_{abcd}C^{abcd}\rightarrow\sqrt{-\tilde{g}} \tilde{C}_{abcd}\tilde{C}^{abcd}
\end{equation}}

Now it is possible to rewrite the action~(\ref{eq:actionJordanfofR}) in the form
{\small 
\begin{eqnarray}
{\cal S}_E &=& \int d^4 x \sqrt{-\tilde{g}}\left[\frac{1}{2 \kappa}\tilde{R}-\frac{1}{2}\tilde{g}^{p q}\partial_p{\phi}\partial_q{\phi}-V(\phi)\right] \nonumber \\
&+& \int d^4 x \mathcal{L}_M(F^{-1}(\phi)\tilde{g}_{pq},\psi_M)\nonumber\\&-&\tcr{\frac{\gamma}{2\kappa}\int d^4x \sqrt{-\tilde{g}} \tilde{C}_{abcd}\tilde{C}^{abcd}}
\end{eqnarray}
}
where 
\begin{equation}
\phi=\sqrt{\frac{3}{2\kappa}} \ln{F} \quad \mbox{and} \quad V(\phi)=\frac{FR-f}{2\kappa F2}.
\end{equation}
It is important to note that there is an additional scalar field in Einstein frame which is directly coupled to matter.
The scalar field equation in the Einstein frame is given by
\begin{equation}
\tilde{\Box}\phi -V_{,\phi} + \sqrt{\kappa} Q\tilde{T}=0
\label{eq:FieldeqnEinstein}
\end{equation}
 where $Q=\frac{-F_{,\phi}}{2\sqrt{\kappa}F}$ quantifies the strength of coupling between the field and matter. \tcr{Since the Weyl part of the action is not coupled to the inflation field, and Eq. \eqref{eq:FieldeqnEinstein} is obtained by varying the action with respect of $\phi$ the Weyl action has no contributions here.}

From Eq.~(\ref{eq:FieldeqnEinstein}), it is clear that $\phi$ is directly coupled to matter.

 Now in a conformal transformation from Jordan frame to Einstein frame the metric transforms as
 \begin{equation}
 d\tilde{s}^2=\Omega^2 ds^2=F(-dt^2 +a^2(dx^2+dy^2+dz^2))
  \end{equation}
  \tcr{In this frame the Bach tensor is zero.Therefore the calculations  presented below hold withoud being changed.}
  Hence we have
  \begin{equation}
  d\tilde{t}=\sqrt{F}dt, \quad \quad \tilde{a}=\sqrt{F} a
  \end{equation}
  Also the Hubble parameter in two frames are related by
  \begin{equation}
  \tilde{H}=\frac{1}{\sqrt{F}} \left(H+\frac{\dot{F}}{2 F}\right)
  \end{equation}
  
  \subsection{Starobinsky Inflation}
  In this section we study a simple inflationary model in $f(R)$ gravity introduced by Starobinsky in 1980~\cite{1980-Starobinsky-PLB}. It is important to note that this is also the first model of inflation. 
  \subsubsection{Dynamics in Jordan frame}
  In this model he considered an action of form 
  \begin{equation}
  f(R)=R+R^2/(6 M^2)
  \label{staromodel}
  \end{equation}
  where $M$ is a new coupling parameter. An exact de Sitter solution was obtained for $f(R) \propto R^2$, further it was shown that the presence of $R$ along with $\alpha R^2$ will provide an exit mechanism.
  
  The approximate solution for the Starobinsky model given by equation~(\ref{staromodel}) is
  \begin{equation}
  H\simeq H_i -\frac{M^2}{6}(t-t_i)\, ; \quad \quad
  a\simeq a_i e^{[H_i(t-t_i)-(M^2/12)(t-t_i)^2]}.
  \end{equation}
   where $H_i$ and $a_i$ are the Hubble parameter and scale factor at the beginning of inflation. 
   
  \tcr{From the above expression, we have 
  $\dot{H} \simeq - M^2/6$ and substituting in the 
 definition of the slow-roll parameter \eqref{def:slowroll}, we have}
   \begin{equation}
   \epsilon_1\simeq \frac{M^2}{6 H^2}
   \end{equation}
   For inflation to occur, $\epsilon$ has to be less than 1 which demands $M \ll H$. The number of e-foldings is given by
   \begin{equation}
   N\simeq \frac{3 H_i^2}{M^2} \simeq \frac{1}{2 \epsilon_1}
   \end{equation}
  The number of e-foldings ($N_*$) after the relevant scales cross the horizon is approximately $55-60$.
   
\subsubsection{Dynamics in Einstein frame}
In Einstein frame, the Starobinsky model action can be written as 
\begin{equation}
{\cal S}_E = \int d^4 x \sqrt{-\tilde{g}}\left[\frac{1}{2 \kappa}\tilde{R}-\frac{1}{2}\tilde{g}^{p q}\partial_p{\phi}\partial_q{\phi}-V(\phi)\right]
\end{equation}
where the field $\phi$ and the potential $V(\phi)$ are given by
\begin{eqnarray}
\phi&=& \sqrt{\frac{3}{2 \kappa}} \ln{F}=\sqrt{\frac{3}{2 \kappa}}\ln\left(1+R/(3 M^2)\right) \\
V(\phi) &=& \frac{3M^2}{4 \kappa}\left(1-2 e^{-\sqrt{2/(3 \kappa)} \phi} \right)
\end{eqnarray}
Following the slow-roll analysis for scalar field, we have
\begin{equation}
\tilde{\epsilon}_1\equiv \frac{d\tilde{H}}{d\tilde{t}}\simeq \frac{1}{2\kappa}\left(\frac{V_{,\phi}}{V}\right)^2 \simeq\frac{4}{3}(e^{\sqrt{2\kappa/3} \phi}-1)^{-2}
\end{equation}
and the number of e-foldings is given by
\begin{equation}
\tilde{N}\simeq \frac{3}{4}e^{\sqrt{2\kappa/3}\phi_i}
\end{equation}